\documentclass[10pt,conference]{IEEEtran}

\usepackage{xspace}
\usepackage{xcolor}
\usepackage{graphicx}
\usepackage[normalem]{ulem} %
\usepackage{amsmath}
\usepackage{amssymb}
\usepackage{amsfonts}
\usepackage{enumitem}
\usepackage{textcomp}

\usepackage{hyperref}
\hypersetup{linkcolor=black,citecolor=black,anchorcolor=black,filecolor=black,menucolor=black,runcolor=black,urlcolor=black,hidelinks}
\usepackage{breakurl}
\usepackage{cite}

\usepackage{booktabs}
\usepackage{multirow}
\usepackage{makecell}
\usepackage{ragged2e}

\usepackage{caption}
\usepackage{subcaption}

\usepackage{listings}
\usepackage{tikz}
\usetikzlibrary{calc}
\usetikzlibrary{shapes.geometric}
\usetikzlibrary{decorations.pathreplacing}
\usetikzlibrary{positioning}

\usepackage{datetime} %

\newcommand{\XSpace}[1]{}
\newcommand{\XComment}[1]{}

\newcommand{\DefMacro}[2]{\expandafter\newcommand\csname rmk-#1\endcsname{#2}}
\newcommand{\UseMacro}[1]{\csname rmk-#1\endcsname}

\newcommand{\InputWithSpace}[1]{\bgroup\def\arraystretch{1.1}\input{#1}\egroup}
\newcommand{\Code}[1]{{\ifmmode{\mathtt{#1}}\else$\mathtt{#1}$\fi}}
\newcommand{\CodeIn}[1]{{\ifmmode{\mathtt{#1}}\else$\mathtt{#1}$\fi}}

\newcolumntype{R}[1]{>{\RaggedLeft\arraybackslash}p{#1}}
\newcolumntype{L}[1]{>{\RaggedRight\arraybackslash}p{#1}}

\definecolor{gray}{RGB}{211,211,211}
\newcommand{\jbasicstyle}{\small\sffamily} %

\newcommand{\jnumberstyle}{\scriptsize}

\lstdefinelanguage{pseudo}
{
morekeywords={},
keywordstyle=\bfseries,
lineskip=-0.1em,
numbers=left, %
numberstyle=\jnumberstyle,
numbersep=4pt,
basicstyle=\jbasicstyle,
breaklines=true,
breakautoindent=true,
tabsize=2,
columns=fullflexible,
morecomment=*[l][\textsl]{//},
mathescape=true,
xleftmargin=10pt,
}

\lstdefinelanguage{todo-comment}
{
morekeywords={},
keywordstyle=\bfseries,
lineskip=-0.1em,
numbers=none,
basicstyle=\jbasicstyle,
breaklines=true,
breakautoindent=true,
tabsize=2,
columns=fullflexible,
morecomment=*[l][\textsl]{//},
mathescape=true,
xleftmargin=-10pt,
}

\lstdefinelanguage{java-pretty}
{
language=java,
numbers=left,
basicstyle=\scriptsize\ttfamily,
numberstyle=\scriptsize,
breaklines=true,
columns=fullflexible,
xleftmargin=16pt,
showstringspaces=false,
}

\newcommand{\Title}{Suggesting Code Edits in Interactive Machine Learning Notebooks Using Large Language Models}

\newcommand{\jupyternotebook}{Jupyter notebook\xspace}
\newcommand{\jupyternotebooks}{Jupyter notebooks\xspace}
\newcommand{\github}{GitHub\xspace}
\newcommand{\revision}{revision\xspace}
\newcommand{\revisions}{revisions\xspace}

\begin{document}

\title{\Title}

\author{
\IEEEauthorblockN{Bihui Jin\textsuperscript{*}}
\IEEEauthorblockA{
{University of Waterloo}\\
Waterloo, Canada \\
bihui.jin@uwaterloo.ca}
\and
\IEEEauthorblockN{Jiayue Wang\textsuperscript{*}}
\IEEEauthorblockA{
{University of Waterloo}\\
Waterloo, Canada \\
jiayue.wang@uwaterloo.ca}
\and
\IEEEauthorblockN{Pengyu Nie}
\IEEEauthorblockA{
{University of Waterloo}\\
Waterloo, Canada \\
pynie@uwaterloo.ca}
}

\maketitle

\begingroup
\renewcommand\thefootnote{\textsuperscript{*}}
\footnotetext{These authors contributed equally to this work.}
\endgroup

\begin{abstract}

Machine learning developers frequently use interactive computational notebooks, such as \jupyternotebooks, to host code for data processing and model training.
\jupyternotebooks provide a convenient tool for writing machine learning pipelines and interactively observing outputs, however, maintaining \jupyternotebooks, e.g., to add new features or fix bugs, can be challenging due to the length and complexity of the notebooks. Moreover, there is no existing benchmark related to developer edits on \jupyternotebooks.
To address this, we present the first dataset of 48,398 \jupyternotebook edits derived from 20,095 \revisions of 792 machine learning repositories on \github, and perform the first study of the using LLMs to predict code edits in \jupyternotebooks.
Our dataset captures granular details of cell-level and line-level modifications, offering a foundation for understanding real-world maintenance patterns in machine learning workflows.
We observed that the edits on \jupyternotebooks are highly localized, with changes averaging only 166 lines of code in repositories.
While larger models outperform smaller counterparts in code editing, all models have low accuracy on our dataset even after finetuning, demonstrating the complexity of real-world machine learning maintenance tasks.
Our findings emphasize the critical role of contextual information in improving model performance and point toward promising avenues for advancing large language models' capabilities in engineering machine learning code.

\end{abstract}

\section{Introduction}
\label{sec:intro}
The widespread adoption of Jupyter notebooks as a development platform in machine learning (ML) has led to its significant role in the iterative processes of coding, debugging, and model training~\cite{Dilhara23PYEVOLVE,Eilertsen20Refactoring,Fluri07TreeDifferencing}.
\jupyternotebooks~\cite{Kluyver2016JupyterNotebooks} are ubiquitous among data scientists and ML developers, serving as synergistic environments for software development~\cite{Ross23SoftwareDevelopmentLLM}, data process~\cite{Wang2023ClinicalGPT}, software evaluation~\cite{Nie23Teco}, and software maintenance~\cite{Zhang23SoftwareMaintenance,Chen23SelfDebug}.
Despite utility, \emph{maintaining} \jupyternotebooks presents significant challenges, such as frequent and fragmented edits, especially as the notebooks grow in size and complexity~\cite{Yee24SMEs}.

The field of software development has been significantly transformed by advancements of large language models (LLMs)~\cite{openai,copilot,Daya24DeepSeek,Baptiste24codeLlama}, offering an alternative for practitioners to efficiently manage repetitive and time-consuming code edits~\cite{Zhiyu22AutomatingCode}. As \jupyternotebooks usually contain domain-specific code on a specific topic, such as machine learning, data-driven approaches like LLMs are also ideal solutions for managing the code edits on them.

Existing work and datasets on \jupyternotebooks~\cite{YinETAL23CodeGenNotebooks,AgasheACL2019JuICe,QuarantaMSR2021KGTorrent,GhahfarokhiMSR2024DistilNotebooks,WangTCHI2021Documentation} focuses exclusively on code generation or comprehension tasks, limiting their relevance for understanding and automating the code maintanance tasks. In addition, they are usually constrained to specific domains (e.g., focusing on speicifc libraries such as pandas and pytorch) and limited in scale.
Therefore, there is a need for a benchmark with realistic ML intents and rich notebook context, so as to better reflect real-world code edits made by ML developers.

To bridge the gap, we propose a dataset with 48,398 \jupyternotebook edits, derived from 20,095 \revisions across 792 machine learning repositories on \github.
We curate a repository dataset with unprecedented granularity, capturing both cell-level and file-level changes alongside associated metadata like commit messages.
Notably, while on average each repository has 8,380 lines of code, the majority of developer edits target specific portions of notebooks, with an average of 166 lines modified per \revision.
By categorizing over 48,000 unique files and 6.63M lines of code, our dataset provides a novel perspective into the evolution of ML code and maintenance patterns in \jupyternotebooks, positioning the dataset as a valuable resource for SE4AI research, particularly in improving the capabilities of LLMs for ML code maintenance.

To demonstrate how our dataset motivates new research on LLMs for ML developers, we explore the utility of LLMs (e.g., DeepSeek~\cite{Daya24DeepSeek}) for automating code edits in \jupyternotebooks on our dataset.
We adopt few-shot prompting strategies to alter the style of model predictions for file-level and cell-level inferences, which improves the diversity of the model's predictions.
Further, we perform a supervised fine-tuning on a large collection of 27,515 \jupyternotebooks in the training set.
Nonetheless, all models have difficulty on our dataset, showing that it is a challenging task.

In summary, the contributions of this work are four-folds:
\begin{itemize}
\item We introduce a dataset of 48,398 \jupyternotebook edits from 20,095 \revisions across 792 machine learning repositories, capturing cell-level and file-level changes, along with commit metadata. This dataset offers unprecedented granularity in tracking developer edits in ML notebooks.

\item Unlike previous datasets, our proposed dataset emphasizes realistic ML development scenarios, featuring detailed insights into localized edits and incremental maintenance practices.

\item As a sample application, we evaluated LLM's performance on our dataset with few-shot learning and supervised fine-tuning, and found that automatically editing \jupyternotebooks is a challenging task for LLMs.

\item We open-source our code for crawling, extracting features, and processing Jupyter notebooks, enabling reproducibility and further research.
\end{itemize}

Our key findings from this study are as follows:

\noindent \textbf{Finding 1: }The state-of-the-art LLMs struggle to effectively handle the dataset's complexity, implying a significant gap between model capabilities and real-world demands.

\noindent \textbf{Finding 2: }Developers typically focus on small, specific sections of notebooks, averaging 166 lines per revision, rather than performing large-scale modifications.

\noindent \textbf{Finding 3: }Providing comprehensive contextual information, such as the full scope of surrounding code, significantly improves LLM performance, particularly for file-level edits.

\noindent \textbf{Finding 4: }While larger models (w/ 6.7B parameters) outperform the lightweight models (w/ 1.3B parameters), neither achieves satisfactory performance on practical development tasks, underscoring the need for advanced fine-tuning techniques and optimization strategies.

\noindent
Our dataset and its replication package are available at\\
\url{https://doi.org/10.5281/zenodo.14281690}

\section{Dataset Construction}
\label{sec:construction}
Figure~\ref{fig:workflow} illustrates an overview of the process of our data collection and processing.
We utilize the Github Search API~\cite{GitHubAPI} to query the top 1000 Github repositories based on the popularity, determined by the number of stars, using the topics ``juypter-notebook'' and ``machine learning''.
After identifying candidate repositories, we clone them and extract the following information:

\begin{figure}
\hspace*{-1.2cm}
\centering
\includegraphics[width=1.2\linewidth]{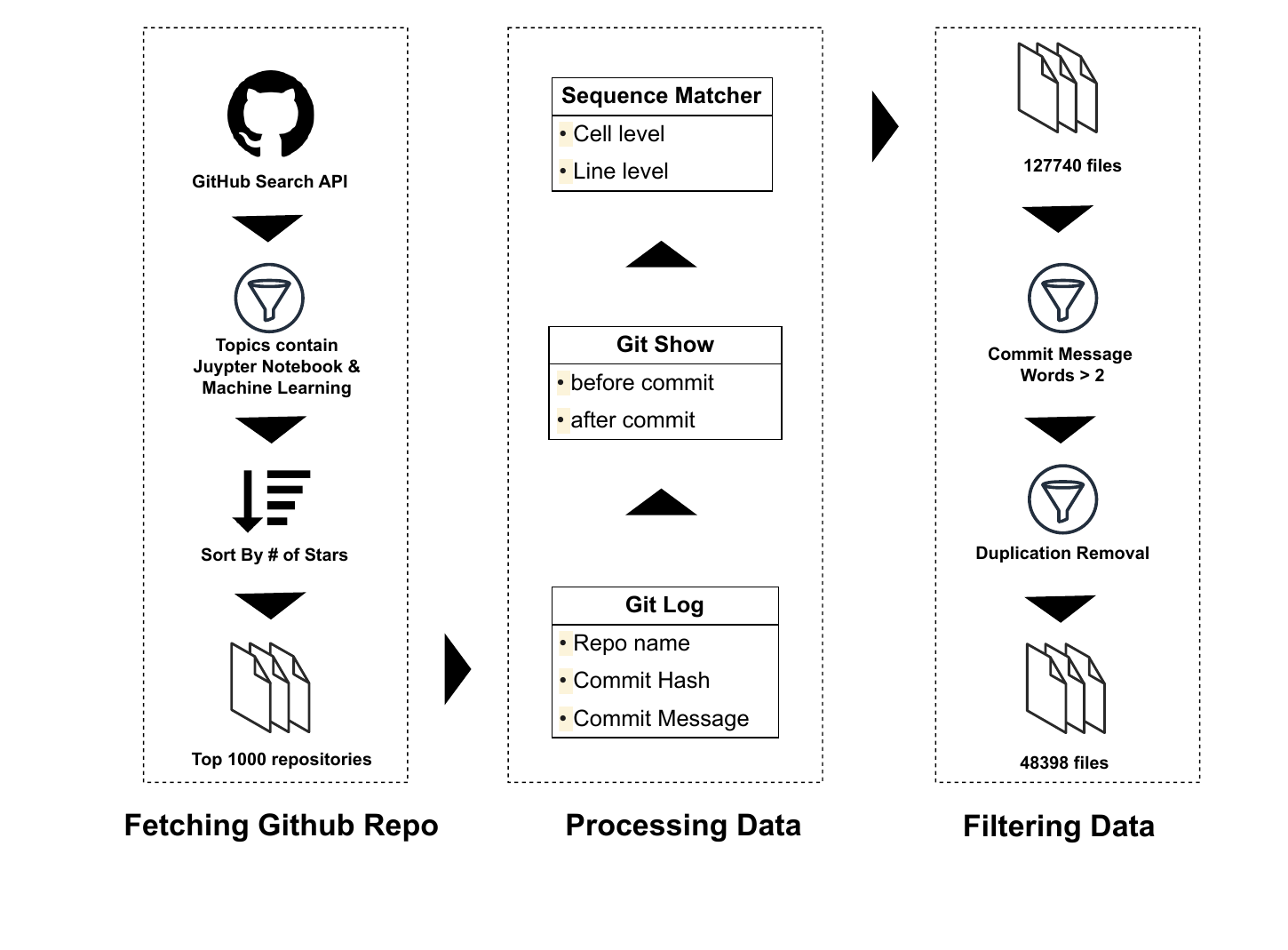}
\vspace*{-1cm}
\caption{An overview of our data collection and processing workflow.}
\label{fig:workflow}
\vspace*{-0.5cm}
\end{figure}

\textbf{1)} We use \textit{git log} with \textit{--name-only} option to retrieve the commit history of each repository.

\textbf{2)} We collect the repository name, commit hash, and commit message for each entry in the git log results that modifies a Juypter Notebook file.

\textbf{3)} For each commit that changes a Juypter Notebook file, we use \textit{git show} to fetch the commit hash along with the content of the code cells in the file, both before and after the commit.

\textbf{4)} Using the \textit{SequenceMatcher} module from \textit{difflib}, we identify cell-level differences between the pre- and post-commit versions. Cells with no changes are omitted.

\textbf{5)} For each cell-level difference with only the modification type rather than additions or deletions, we compare line-level difference.
Lines with no changes are omitted, which results in
a total of 127,740 file-level edits, including line-level and cell-level edits.
The collected information includes the repository name, commit hash, file name, commit message, content before the commit, content after the commit, cell level differences, and line level differences.

\textbf{6)} We refine the dataset by excluding commit messages containing two or fewer words and removing duplicate entries.

As a result, we obtain 48,398 files, leaving the final dataset with meaningful and unique commit information selected from the repositories.

\section{Dataset Description And Statistics}
\label{sec:dataset}

Our dataset captures detailed insights into the evolution of Jupyter notebooks within ML repositories, offering a granular perspective on their maintenance and editing practices.

\textbf{Comparing with Existing Datasets.}
Existing datasets of \jupyternotebooks mined from open-source repositories either focus on code generation~\cite{YinETAL23CodeGenNotebooks,AgasheACL2019JuICe}, code comprehension~\cite{WangTCHI2021Documentation} or metadata and code metrics~\cite{QuarantaMSR2021KGTorrent,GhahfarokhiMSR2024DistilNotebooks}, and lack granularity in tracking developers' code editing behaviors or \revisions over time.
They are constrained to specific domains (e.g., focusing on the usages of specific libraries) and limited in scale.
Our dataset is the first one of code edits in \jupyternotebooks and covers a wide range of repositories on machine learning topic.

\begin{table}[t]
\begin{scriptsize}
\centering
\caption{Statistics of our dataset.}\label{tab:statistics}

\resizebox{\linewidth}{!}{%
\begin{tabular}{l|r rrrrr}
\toprule
Name & Avg & Min & Med & 75\% & 90\% & Max \\
\toprule
commit message \#Words & 5.66 & 3 & 5 & 7 & 9 & 72 \\
file \#Token before diff & 1,511.60 & 1 & 593 & 2,092 & 4,191 & 50,451\\
file \#Token after diff & 2,079.28 & 1 & 1,260 & 2,756 & 4,845 & 79,893\\  %
changed cells \#Token before diff & 344.14 & 1 & 1 & 204 & 944 & 30,800\\  %
changed cells \#Token after diff & 841.14 & 1 & 191 & 947 & 2,437 & 77,597\\  %
\bottomrule
\end{tabular}
}

\end{scriptsize}
\end{table}

\textbf{Individual commits recorded in \jupyternotebooks are focused on specific areas of the notebook rather than making changes to the entire notebook.}
Table~\ref{tab:statistics} outlines key statistical features of our dataset, including numbers of words in commit messages as well as token counts in files and cells before and after commits, reflecting the scale and focus of revisions in computational notebooks.
The observed median values for file token counts and changed cell sizes indicate that even within large repositories, the majority of revisions are incremental rather than major modification that affect large portions of the repository, such as restructuring entire files, refactoring major components, and rewriting sections of code.

On average, commit messages are concise, comprising 5.66 words, with 75\% containing fewer than 7 words, indicating that developers prioritize brevity when describing changes.
The files undergoing edits show an average of 1,511.60 tokens before the commit, increasing to 2,079.28 tokens post-edit, which suggests a focus on incremental development and the inclusion of additional content during updates.
Interestingly, the tokens in modified cells highlight the localized edits in which developers often target specific sections with an average of 344.14 tokens pre-edit and 841.14 tokens post-edit, which emphasizes the limited scope of most changes.

\begin{table}[t]
\centering
\caption{Size of our dataset after splitting.}\label{tab:split}

\resizebox{\linewidth}{!}{%
\begin{tabular}{l|rrrr|rr}
\toprule
Set   & \#Proj & \#File & LOC     & \#Commits & \#$\Delta$Cells & \#$\Delta$Lines \\ \toprule
full  & 792    & 48,398  & 6,636,971 & 20,095     & 508,634                          & 3,336,142                         \\ \hline
train & 536    & 27,515  & 3,510,884 & 12,874     & 308,731                          & 1,899,031                         \\
val   & 76     & 12,469  & 1,988,162 & 3,660      & 116,313                          & 970,452                          \\
test  & 152    & 8,414   & 1,137,925 & 3,561      & 83,590                           & 466,659                          \\ \hline
\end{tabular}
}

\end{table}

\textbf{Developers typically address targeted issues or implement small updates without disrupting the broader codebase. }
Table~\ref{tab:split} illustrates a breakdown of the dataset’s scale and structure, categorizing it into training, validation, and test subsets.
The majority of revisions focus on localized editing, where changes are confined to specific cells or segments within a notebook, rather than overhauling entire notebooks or files.
The full dataset spans 792 projects (\#Proj), 48,398 files (\#File), and over 6.6 million lines of code (LOC).
Of the 20,095 commits, changes are localized, with approximately 508,634 cell edits (\#$\Delta$Cells) and 3,336,142 line modifications (\#$\Delta$Lines), reinforcing the observation that developers typically make focused updates rather than wholesale changes.
The dataset splits into 536 repositories for training, 76 for validation, and 152 for testing, for the evaluation in Section~\ref{sec:eval}.
The training set, comprising 27,515 files and 3.5 million LOC, represents the majority of the dataset, while the test set maintains a significant size with 8,414 files and 1.1 million LOC, supporting diverse evaluation scenarios.

\section{Evaluation}
\label{sec:eval}
We explore two inference strategies: few-shot prompting and fine-tuning with few-shot exemplars to impose more control on the predicted code’s style.

\subsection{Models}
DeepSeek-Coder~\cite{Daya24DeepSeek} is a pre-trained decoder-only transformer model tailored for code-related tasks, utilizing a 16,384-token context window and a fill-in-the-blank task to enhance code generation and infilling capabilities.
We employ the instructed DeepSeek-Coder models with 1.3 billion parameters (i.e., DeepSeek-Coder-1.3b-instruct) and 6.7 billion parameters (i.e., DeepSeek-Coder-6.7b-instruct) to evaluate abilities of inferring file-level and cell-level code changes.
The model is tasked with predicting the code edits with five-shot exemplars.

\subsection{Few-shot Prompting}
We provide a system prompt that assigns the model the role of an expert software developer.
For each code edit task, we create specific prompts with five random examples (five-shot) from training set tailored to the type of edits being performed:

\textbf{File-level prompts.}
The prompt includes instructions derived from the commit messages and the content of the code cells before the commit.
Additionally, we provide a sample response using the actual content of the code cells after the commit.

\textbf{Cell-level prompts.}
The prompt consists of instructions based on the commit messages and the concatenation of modified code cells made before the commit, where the sample response is the concatenation of code cells after the commit.

A ``\#Response" instruction with the committed code edits is included to guide the model in addressing either file-level or cell-level code edits while leveraging a few-shot learning strategy to improve its performance.
For examples from the test set, the model is expected to generate responses based solely on the provided instructions.

\subsection{Evaluation Metrics}
The following automatic metrics are employed to measure the similarity between the predicted statements and the reference (gold) statements. These metrics are widely utilized in previous studies on code generation~\cite{Zhang23Coevolution,Nie23Teco}.

\textbf{BLEU}~\cite{Papineni02BLEU} evaluates the overlap of 1 $\sim$ 4-grams (sequences of 1-4 consecutive subtokens) between the predicted output and the gold, averaging the results across these 1 $\sim$ 4-grams using a smoothing technique introduced by Lin and Och~\cite{Lin04ORANGE}.

\textbf{CodeBLEU}~\cite{Ren20CodeBLEU} extends BLEU for code-specific tasks by incorporating additional dimensions: traditional BLEU, BLEU focusing solely on keywords, syntactic matching through abstract syntax trees, and semantic data-flow match.

\textbf{Edit similarity} (EditSim)~\cite{Svyatkovskiy20IntelliCode} is defined as 1 $-$ Levenshtein edit distance, where the Levenshtein edit distance quantifies the minimum number of single-character edits, including insertions, substitutions, or deletions, which require to transform the prediction into the gold, normalized by the maximum length of characters in the prediction and the gold.

\textbf{RougeL}~\cite{Lin2004Rouge} uses the F1 score to assess the overlap between the prediction and reference subtokens by analyzing the longest common subsequence statistics.

\subsection{Fine-tuning}
We then fine-tune the basic DeepSeek-Coder model on a large collection of approximate 13K revisions collected from 536 GitHub repositories.
We fine-tune three DeepSeek-Coder instructed models, each with 1.3 billion parameters, using zero-shot exemplars for each cell-level and file-level tasks.
Besides the basic setting, we also experiment with advanced techniques, such as LoRA~\cite{Hu22lora}, PiSSA~\cite{Meng2024pissa}, and FlashAttention-v2~\cite{Dao23flashattention2} to further enhance the the model's robustness and efficiency.
To evaluate the impact of fine-tuning, we compare the performance of the pre-trained DeepSeek-Coder models (without fine-tuning) against the average results of the three fine-tuned models across four evaluation metrics for distinct tasks.

\subsection{Results}

Table~\ref{tab:results-cell} and Table~\ref{tab:results-file} compare the performance of DeepSeek-Coder models on the code edit task, focusing on localized cell-level edits and global file-level edits, based on developer requirements.

\begin{table}[t]
\centering
\caption{Results of predicting code changes within the changed cells.}
\label{tab:results-cell}

\resizebox{\linewidth}{!}{%
\begin{tabular}{l|rrrr}
\hline
DeepSeek Model & BLEU  & CodeBLEU & EditSim & RougeL \\ \hline
1.3B           & 8.52 & 17.26    & 26.52   & 21.49  \\
6.7B           & 13.30 & 20.49    & 31.01   & 28.05  \\
1.3B-postproc  & 8.59 & 18.12    & 26.64   & 21.66  \\
6.7B-postproc  & 13.44 & 22.06    & 31.23   & 28.30  \\ \hline
1.3B-finetune  & \textbf{25.86} & \textbf{30.55}    & \textbf{46.53}   & \textbf{40.80}  \\ \hline
\end{tabular}
}

\end{table}

\textbf{All models face difficulties on our dataset, reflecting its complexity, which subsumes challenging ML tasks and real-world practical issues.}
Table~\ref{tab:results-cell} displays the performance of the DeepSeek models on cell-level code change predictions.
Specifically, the BLEU scores for the 1.3B model and the 6.7B model with five-shot exemplars are only 8.521 and 13.30, respectively,
Similarly, the CodeBLEU scores are 17.26 and 20.49, while EditSim scores reach 26.52 and 31.01, respectively.
Although the 6.7B model performs marginally better than the 1.3B model, the low BLEU scores (both below 19) imply that the predictions lack clarity and are of limited utility.
Post-processing the inference results by removing code block symbols (\verb|```|) and removing generated instruction prompts (e.g., commit messages) improves the scores.
For the 1.3B model, post-processing increases performance by 0.45\% to 4.98\%, depending on the metric, while the 6.7B model makes improvements of 0.71\% to 7.66\%.

\begin{table}[t]
\centering
\caption{Results of predicting code changes at file level.}
\label{tab:results-file}

\resizebox{\linewidth}{!}{%
\begin{tabular}{l|rrrr}
\hline
DeepSeek Model & BLEU  & CodeBLEU & EditSim & RougeL \\ \hline
1.3B           & 10.75 & 15.39    & 31.78   & 27.23  \\
6.7B           & 11.76 & 15.34    & 34.68   & 30.45  \\
1.3B-postproc  & 10.78 & 17.42    & 31.76   & 27.36  \\
6.7B-postproc  & 11.81 & 18.64    & 34.70   & 30.57  \\ \hline
1.3B-finetune  & \textbf{27.46} & \textbf{45.26}    & \textbf{47.38}   & \textbf{53.13}  \\ \hline
\end{tabular}
}

\end{table}

\textbf{While the 6.7B model consistently outperforms the 1.3B model, the improvements due to post-processing still remain modest, indicating limitations in the models’ inherent reasoning capabilities for file-level edits.}
Table~\ref{tab:results-file} provides the performance of the DeepSeek models on file-level code change predictions.
Overall, The 6.7B model slightly outperforms the 1.3B model, achieving a BLEU score of 11.76 compared to 10.75.
Such trend is consistent across other metrics, with the 6.7B model recording a CodeBLEU of 15.34, EditSim of 34.68, and ROUGEL of 30.45, compared to the 1.3B model’s CodeBLEU of 15.39, EditSim of 31.78, and ROUGEL of 27.23.
Post-processing also yields further improvements on the file-level edits.
For the 1.3B model, post-processing increases performance by -0.06\% to 13.19\%, depending on the metric, while the 6.7B model makes improvements of 0.06\% to 21.51\%.

\textbf{While larger models, such as the 6.7B model, consistently outperform the smaller counterparts (i.e., 1.3B model), neither meets the demands of real-world development tasks.}
Table~\ref{tab:results-cell} and Table~\ref{tab:results-file} indicate that the DeepSeek-Coder-6.7b-instruct model is the best model in terms of BLEU, CodeBLEU, EditSim and ROUGE metrics.
Nevertheless, lightweight LLMs still fall short for practical use.
Although larger models like the 6.7B perform better, their limitations highlight the need for supplementary techniques to enhance reasoning capabilities and further advancements in model intelligence to tackle with real-world challenges effectively.

\textbf{LLMs demonstrate superior performance on tasks like file-level edits when provided with more comprehensive contextual information, such as the full scope of the surrounding code.}
Although the finetuned 1.3B model attains a remarkable improvement in performance (delivering BLEU scores around 3x higher than the 1.3B model and twice as high as the 6.7B model, while also doubling RougeL scores), the persistence of syntax errors underscores that \textbf{our dataset is challenging in practical ML editing tasks.}
The substantial gains in BLEU and ROUGE-L suggest that the finetuned model could be viable for applications where capturing the overall gist is sufficient.
While the finetuned model excels in inferring clear gist, the finetuned model demonstrates the critical importance of task-specific optimization for improving both syntactical and semantic performance.

\section{Conclusions and Future Work}
\label{sec:conclusion}

We presented the first dataset for studying code edits in \jupyternotebooks, consisting of 48,398 edits from 20,095 \revisions of 792 machine learning repositories mined from on \github. Different from existing datasets of \jupyternotebooks mined from \github, ours provide a foundation for understanding real-world maintenance patterns in machine learning workflows. In our dataset, we preserved developer edits on at both cell-level and line-level, so that future work can analyze and leverage our dataset in a flexible way. We also documented our dataset construction process in details and open-sourced our dataset and collection scripts to allow replication.

As a demonstration, we performed a set of experiments of applying LLMs to predict \jupyternotebook edits on our dataset. We tried both few-shot learning and supervised fine-tuning, as well as applying LLMs on two different granularities: predicting the entire file and predicting only the changed cells.
Interestingly, all models exhibit low accuracy on our dataset, even after fine-tuning, highlighting the complexity of real-world machine learning maintenance tasks.

Based on our experience, we plan to explore retrieval-augmented-generation and agentic techniques to improve the performance of LLMs on predicting \jupyternotebook edits, as extracting relevant contextual information is critical in performing targeted edits in usually lengthy notebooks.

\newpage
\bibliography{bib}

\begin{thebibliography}{10}
\providecommand{\url}[1]{#1}
\csname url@samestyle\endcsname
\providecommand{\newblock}{\relax}
\providecommand{\bibinfo}[2]{#2}
\providecommand{\BIBentrySTDinterwordspacing}{\spaceskip=0pt\relax}
\providecommand{\BIBentryALTinterwordstretchfactor}{4}
\providecommand{\BIBentryALTinterwordspacing}{\spaceskip=\fontdimen2\font plus
\BIBentryALTinterwordstretchfactor\fontdimen3\font minus
  \fontdimen4\font\relax}
\providecommand{\BIBforeignlanguage}[2]{{%
\expandafter\ifx\csname l@#1\endcsname\relax
\typeout{** WARNING: IEEEtranS.bst: No hyphenation pattern has been}%
\typeout{** loaded for the language `#1'. Using the pattern for}%
\typeout{** the default language instead.}%
\else
\language=\csname l@#1\endcsname
\fi
#2}}
\providecommand{\BIBdecl}{\relax}
\BIBdecl

\bibitem{Fluri07TreeDifferencing}
``Change distilling:tree differencing for fine-grained source code change
  extraction,'' \emph{IEEE Transactions on Software Engineering}, vol.~33,
  no.~11, pp. 725--743, 2007.

\bibitem{AgasheACL2019JuICe}
R.~Agashe, S.~Iyer, and L.~Zettlemoyer, ``Juice: A large scale distantly
  supervised dataset for open domain context-based code generation,'' in
  \emph{Conference on Empirical Methods in Natural Language Processing}, 2019.

\bibitem{Chen23SelfDebug}
X.~Chen, M.~Lin, N.~Sch{\"a}rli, and D.~Zhou, ``Teaching large language models
  to self-debug,'' \emph{ArXiv}, vol. abs/2304.05128, 2023.

\bibitem{Dao23flashattention2}
T.~Dao, ``Flashattention-2: Faster attention with better parallelism and work
  partitioning,'' 2023.

\bibitem{Dilhara23PYEVOLVE}
M.~Dilhara, D.~Dig, and A.~Ketkar, ``Pyevolve: Automating frequent code changes
  in python ml systems,'' in \emph{2023 IEEE/ACM 45th International Conference
  on Software Engineering (ICSE)}, 2023, pp. 995--1007.

\bibitem{Eilertsen20Refactoring}
A.~M. Eilertsen, ``Refactoring operations grounded in manual code changes,'' in
  \emph{Proceedings of the ACM/IEEE 42nd International Conference on Software
  Engineering: Companion Proceedings}, ser. ICSE '20.\hskip 1em plus 0.5em
  minus 0.4em\relax New York, NY, USA: Association for Computing Machinery,
  2020, p. 182–185.

\bibitem{GhahfarokhiMSR2024DistilNotebooks}
M.~M. Ghahfarokhi, A.~Asgari, M.~Abolnejadian, and A.~Heydarnoori,
  ``Distilkaggle: A distilled dataset of kaggle jupyter notebooks,'' \emph{2024
  IEEE/ACM 21st International Conference on Mining Software Repositories
  (MSR)}, pp. 647--651, 2024.

\bibitem{GitHubAPI}
\BIBentryALTinterwordspacing
GitHub, ``Rest api endpoints for repositories,'' November 2022. [Online].
  Available: \url{https://docs.github.com/en/rest/repos?apiVersion=2022-11-28}
\BIBentrySTDinterwordspacing

\bibitem{Daya24DeepSeek}
D.~Guo, Q.~Zhu, D.~Yang, Z.~Xie, K.~Dong, W.~Zhang, G.~Chen, X.~Bi, Y.~Wu,
  Y.~Li, F.~Luo, Y.~Xiong, and W.~Liang, ``Deepseek-coder: When the large
  language model meets programming -- the rise of code intelligence,'' 2024.

\bibitem{Hu22lora}
\BIBentryALTinterwordspacing
E.~J. Hu, Y.~Shen, P.~Wallis, Z.~Allen-Zhu, Y.~Li, S.~Wang, L.~Wang, and
  W.~Chen, ``Lo{RA}: Low-rank adaptation of large language models,'' in
  \emph{International Conference on Learning Representations}, 2022. [Online].
  Available: \url{https://openreview.net/forum?id=nZeVKeeFYf9}
\BIBentrySTDinterwordspacing

\bibitem{Kluyver2016JupyterNotebooks}
T.~Kluyver, B.~Ragan-Kelley, F.~P{\'e}rez, B.~E. Granger, M.~Bussonnier,
  J.~Frederic, K.~Kelley, J.~B. Hamrick, J.~Grout, S.~Corlay, P.~Ivanov,
  D.~Avila, S.~Abdalla, C.~Willing, and J.~D. Team, ``Jupyter notebooks - a
  publishing format for reproducible computational workflows,'' in
  \emph{International Conference on Electronic Publishing}, 2016.

\bibitem{Zhiyu22AutomatingCode}
Z.~Li, S.~Lu, D.~Guo, N.~Duan, S.~Jannu, G.~Jenks, D.~Majumder, J.~Green,
  A.~Svyatkovskiy, S.~Fu, and N.~Sundaresan, ``Automating code review
  activities by large-scale pre-training,'' in \emph{Proceedings of the 30th
  ACM Joint European Software Engineering Conference and Symposium on the
  Foundations of Software Engineering}, ser. ESEC/FSE 2022.\hskip 1em plus
  0.5em minus 0.4em\relax New York, NY, USA: Association for Computing
  Machinery, 2022, p. 1035–1047.

\bibitem{Lin2004Rouge}
C.-Y. Lin and F.~J. Och, ``Automatic evaluation of machine translation quality
  using longest common subsequence and skip-bigram statistics,'' in
  \emph{Annual Meeting of the Association for Computational Linguistics}, 2004.

\bibitem{Lin04ORANGE}
------, ``Orange: a method for evaluating automatic evaluation metrics for
  machine translation,'' in \emph{International Conference on Computational
  Linguistics}, 2004.

\bibitem{Meng2024pissa}
\BIBentryALTinterwordspacing
F.~Meng, Z.~Wang, and M.~Zhang, ``Pissa: Principal singular values and singular
  vectors adaptation of large language models,'' 2024. [Online]. Available:
  \url{https://arxiv.org/abs/2404.02948}
\BIBentrySTDinterwordspacing

\bibitem{copilot}
\BIBentryALTinterwordspacing
Microsoft, ``Search microsoft copilot: Your everyday ai companion.'' [Online].
  Available: \url{https://copilot.microsoft.com/}
\BIBentrySTDinterwordspacing

\bibitem{Nie23Teco}
P.~Nie, R.~Banerjee, J.~J. Li, R.~J. Mooney, and M.~Gligori{\'c}, ``Learning
  deep semantics for test completion,'' \emph{2023 IEEE/ACM 45th International
  Conference on Software Engineering (ICSE)}, pp. 2111--2123, 2023.

\bibitem{openai}
\BIBentryALTinterwordspacing
OpenAI, ``Gpt-4 technical report,'' 2024. [Online]. Available:
  \url{https://arxiv.org/abs/2303.08774}
\BIBentrySTDinterwordspacing

\bibitem{Papineni02BLEU}
K.~Papineni, S.~Roukos, T.~Ward, and W.-J. Zhu, ``Bleu: a method for automatic
  evaluation of machine translation,'' in \emph{Annual Meeting of the
  Association for Computational Linguistics}, 2002.

\bibitem{QuarantaMSR2021KGTorrent}
L.~Quaranta, F.~Calefato, and F.~Lanubile, ``Kgtorrent: A dataset of python
  jupyter notebooks from kaggle,'' \emph{2021 IEEE/ACM 18th International
  Conference on Mining Software Repositories (MSR)}, pp. 550--554, 2021.

\bibitem{Ren20CodeBLEU}
S.~Ren, D.~Guo, S.~Lu, L.~Zhou, S.~Liu, D.~Tang, M.~Zhou, A.~Blanco, and S.~Ma,
  ``Codebleu: a method for automatic evaluation of code synthesis,''
  \emph{ArXiv}, vol. abs/2009.10297, 2020.

\bibitem{Ross23SoftwareDevelopmentLLM}
S.~I. Ross, F.~Martinez, S.~Houde, M.~J. Muller, and J.~D. Weisz, ``The
  programmer’s assistant: Conversational interaction with a large language
  model for software development,'' \emph{Proceedings of the 28th International
  Conference on Intelligent User Interfaces}, 2023.

\bibitem{Baptiste24codeLlama}
B.~Rozière, J.~Gehring, F.~Gloeckle, S.~Sootla, I.~Gat, X.~E. Tan, Y.~Adi,
  J.~Liu, R.~Sauvestre, T.~Remez, J.~Rapin, A.~Kozhevnikov, I.~Evtimov,
  J.~Bitton, M.~Bhatt, C.~C. Ferrer, A.~Grattafiori, W.~Xiong, A.~Défossez,
  J.~Copet, F.~Azhar, H.~Touvron, L.~Martin, N.~Usunier, T.~Scialom, and
  G.~Synnaeve, ``Code llama: Open foundation models for code,'' 2024.

\bibitem{Svyatkovskiy20IntelliCode}
A.~Svyatkovskiy, S.~K. Deng, S.~Fu, and N.~Sundaresan, ``Intellicode compose:
  code generation using transformer,'' \emph{Proceedings of the 28th ACM Joint
  Meeting on European Software Engineering Conference and Symposium on the
  Foundations of Software Engineering}, 2020.

\bibitem{WangTCHI2021Documentation}
A.~Y. Wang, D.~Wang, J.~Drozdal, M.~J. Muller, S.~Park, J.~D. Weisz, X.~Liu,
  L.~Wu, and C.~Dugan, ``Documentation matters: Human-centered ai system to
  assist data science code documentation in computational notebooks,''
  \emph{ACM Transactions on Computer-Human Interaction}, vol.~29, pp. 1 -- 33,
  2021.

\bibitem{Wang2023ClinicalGPT}
G.~Wang, G.~Yang, Z.~Du, L.~Fan, and X.~Li, ``Clinicalgpt: Large language
  models finetuned with diverse medical data and comprehensive evaluation,''
  \emph{ArXiv}, vol. abs/2306.09968, 2023.

\bibitem{Yee24SMEs}
J.~S.~G. Yee, P.~C. Ng, Z.~Wang, I.~McLoughlin, A.~B. Ng, and S.~See,
  ``On-device llms for smes: Challenges and opportunities,'' \emph{ArXiv}, vol.
  abs/2410.16070, 2024.

\bibitem{YinETAL23CodeGenNotebooks}
P.~Yin, W.-D. Li, K.~Xiao, A.~Rao, Y.~Wen, K.~Shi, J.~Howland, P.~Bailey,
  M.~Catasta, H.~Michalewski, O.~Polozov, and C.~Sutton, ``Natural language to
  code generation in interactive data science notebooks,'' in \emph{Annual
  Meeting of the Association for Computational Linguistics}, 2023, pp.
  126--173.

\bibitem{Zhang23Coevolution}
J.~Zhang, P.~Nie, J.~J. Li, and M.~Gligori{\'c}, ``Multilingual code
  co-evolution using large language models,'' \emph{Proceedings of the 31st ACM
  Joint European Software Engineering Conference and Symposium on the
  Foundations of Software Engineering}, 2023.

\bibitem{Zhang23SoftwareMaintenance}
Y.~Zhang, ``Large language model in sd-wan intelligent operations and
  maintenance,'' \emph{Research Briefs on Information and Communication
  Technology Evolution}, 2023.

\end{thebibliography}

\end{document}